# MIMAC-HE3: A NEW DETECTOR FOR NON-BARYONIC DARK MATTER SEARCH

DANIEL SANTOS

*Laboratoire de Physique Subatomique et de Cosmologie (CNRS-IN2P3/Université Joseph Fourier) 53, Av. des Martyrs*
*38026 Grenoble, France*

EMMANUEL MOULIN

*Laboratoire de Physique Subatomique et de Cosmologie (CNRS-IN2P3/Université Joseph Fourier) 53, Av. des Martyrs*
*38026 Grenoble, France*

The project of a micro-TPC matrix of cells of 3He for direct detection of non-baryonic dark matter is presented. The privileged properties of 3He for this detection are highlighted. The double detection: ionization – projection of tracks is explained and its rejection evaluated. The specific capabilities of this project with respect to other experiments are mentioned and its complementarities concerning the supersymmetric phenomenology explicitly showed.

## 1. Why 3He ?

In the last years, our work on $^3$He as a target for detecting WIMPs allowed us to confirm its privileged properties for direct detection [1].
These properties can be enumerated as follow: i) its fermionic character opens the axial interaction with fermionic WIMPs, as the neutralinos, ii) the extremely low Compton cross section reduces in several order of magnitude the natural radioactive background with respect to other targets, iii) the high neutron capture cross section gives a clear signature for neutron rejection, iv) its light mass allows a higher sensitivity to light WIMP masses than other targets, v) the elastic energy transfer is bounded to a very narrow range of energy (a few keV) allowing to have all the interesting events concentrated in a narrow energy range offering a high signal to noise ratio.
The extremely low Compton cross section and the possibility to detect events in the keV range (< 5.6 keV) have been demonstrated by the electron conversion 57Co detection recently reported [2]. The fact that electrons of 7 keV could be detected in the MACHe3 [3] prototype with the source emitting the 122 keV gamma rays embedded in the $^3$He is a clear demonstration of the virtual transparency of this medium to the electromagnetic radiation.
The work developed on the MACHe3 prototype concerning the simulations of the interaction of the electromagnetic radiation and cosmic particles and the rejection estimation based on the properties above mentioned [3,4] are applied to a new TPC detector offering the electron – recoil discrimination.

## 2. Micro- TPC

The micro temporal projection chambers with an avalanche amplification using a pixelized anode [5] present the required features to discriminate electron – recoil events by the double detection of the ionization energy and the track projection onto the anode. A schematic structure of the chamber is shown on fig1. In order to get the electron-recoil discrimination, the pressure of the TPC should be such that the tracks of the electrons having energies less than 6 keV could be well resolved with respect to those corresponding to the recoils of the same energy convoluted by the quenching factor.

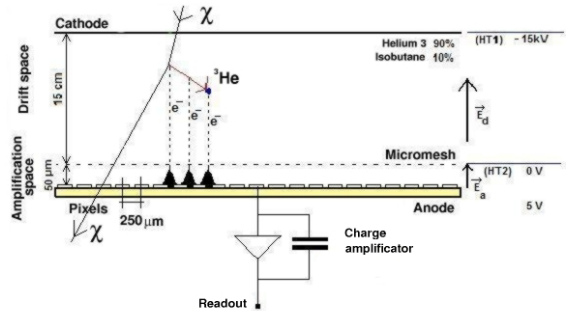

Figure 1. A typical cell chamber of MIMAC-He3 filled with gaseous $^3$He. The pixellised anode allows a 2D projection of the recoil track.

Simulations have been done, as a function of the pressure, for electrons using Geant 4 [6] and for recoils using SRIM [7]. The results are shown on fig 2. The electrons produced by the primary interactions will drift to the grid in a diffusion process following the well known distribution characterized by a radius of D~200μm √ L[cm] where L is the total drift in the chamber up to the grid. This process has been simulated with Garfield [8] and the drift velocities estimated as a function of the pressure and the electric field. A typical value of 26μm/ns is obtained for 1kV/cm at a pressure of 1 bar.

To prevent confusion between the electron tracks projection and the recoil ones the total drift length should be limited to L~15 cm. This fact defines the elementary cell of the detector matrix and the simulations performed on the ranges of electrons and recoils suggest that with an anode of 250 μm the electron-recoil discrimination required can be obtained.

As mentioned above, the quenching factor is an important point that should be addressed to quantify the amount of the total recoil energy recovered in the ionization channel. No measurements of the quenching factor (QF) in $^3$He have been reported. However, an estimation can be obtained applying the Lindhard calculations [9]. We have plotted on fig. 3 the estimated quenching factors for different pairs of nuclei ($^{132}$Xe, $^{74}$Ge, and $^3$He) of the same atomic and mass numbers as a function of the recoil energy. The Ge curve has been validated by a neutron induced measurement more than ten years ago [10]. The $^3$He curve shows that we can expect up to 60% of the recoil energy in the range of interest (<6 keV) going through ionization.

In order to measure the QF in $^3$He at such low energies, we have designed, at the LPSC laboratory, an ion source to accelerate the helium ions to be coupled to the micro-TPC chamber. The ions accelerated by the source will pass through a thin foil of polypropylene that will neutralize them before enter to the chamber. The measurement of the energy of the atoms of $^3$He entering to the chamber will be made by a time of flight measurement. The determination of the QF is one important step of the project and it will be performed shortly.

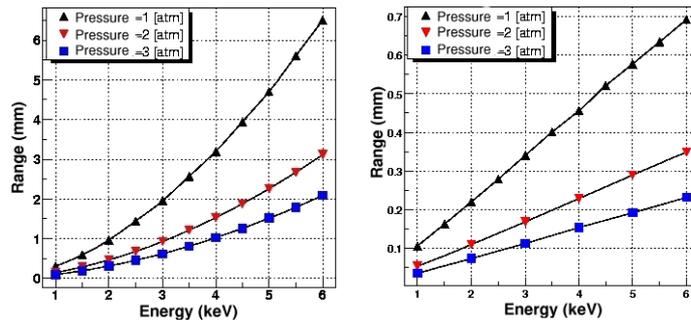

Figure 2. The left (resp. right) side plot shows the electron (resp. $^3$He) simulated range versus the kinetic energy for 1, 2 and 3 atm pressure.

### 3. Double detection: Ionization-Track projection

In order to characterize the distribution of pixels on the anode produced by the different kind of trajectories we define the ratio between perpendicular symmetry axis of the pixel distribution (a/b) where a is the larger axis of the distribution.
We plot on fig. 4 the histograms of the ratio a/b distribution for the simulations performed at different energies of electrons and recoils at different pressures. An isotropic spherical emission of electrons and recoils at L~10 cm from the grid has been injected as the input of the simulation. For the recoils a very concentrated distribution around 1 is expected, and for the electrons a very wide one.
The rejection of events using the a/b ratio is a strong function of the energy and the pressure of the chamber, but even at 1 keV and 3 bar only a small number of the total events can be confused .

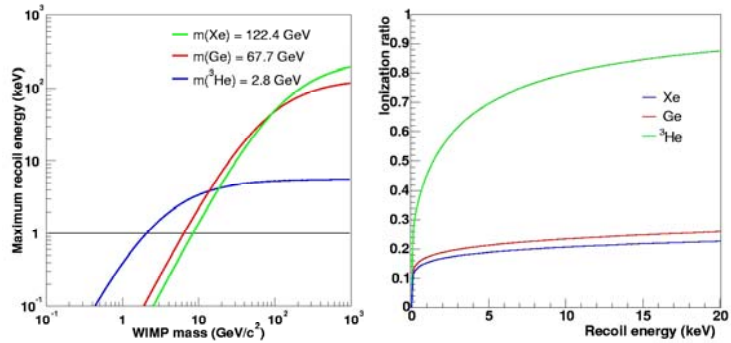

Figure 2 . The plot on the left shows the maximum recoil energy of the target nucleus versus the incident WIMP mass for different target nuclei. In the case of $^3$He, the energy range where all the searched events have to fall in is from the energy threshold up to 5.6 keV. For heavier nuclei, the energy range is much wider. The plot on the right side shows the ionization ratio predicted by Lindhard [9] for different sensitive media. For $^3$He, up to 60% of the recoil energy is expected to be released in the ionization channel.

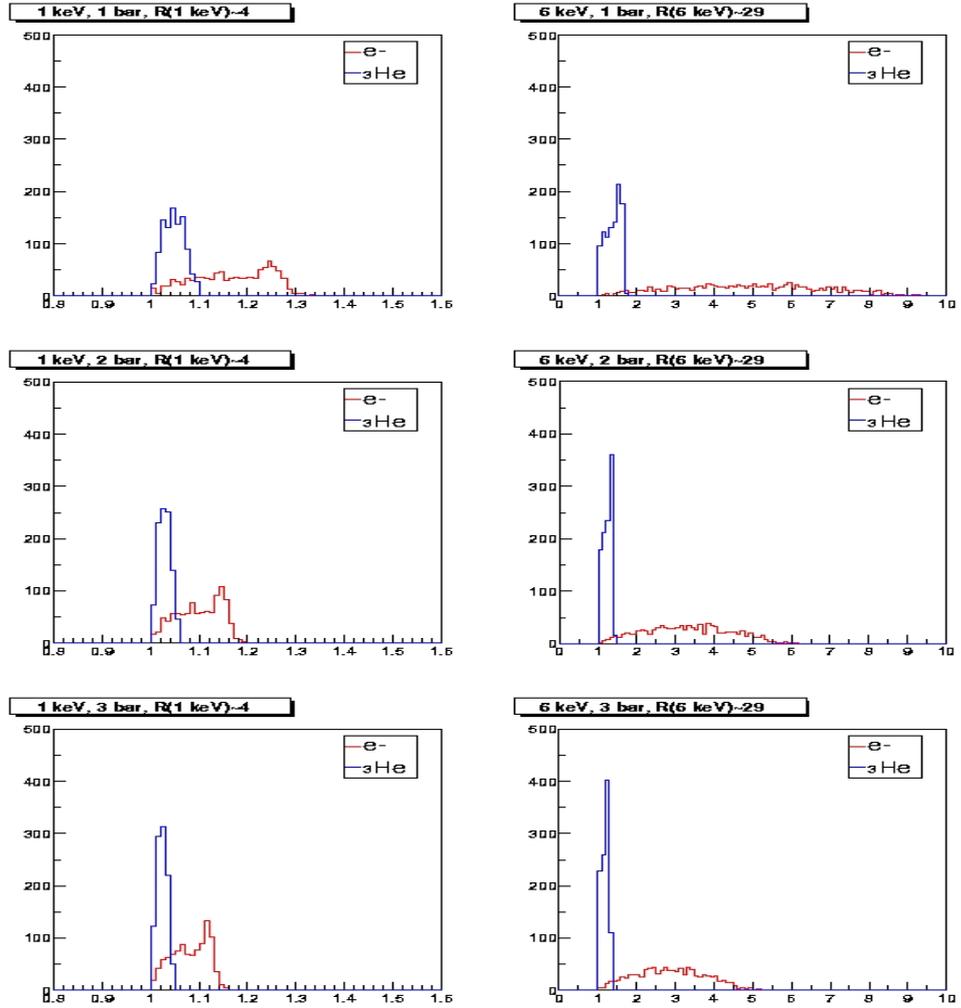

Figure 4. The histograms of the a/b distribution, for a recoil energy of 1 and 6 keV, and a pressure of 1, 2 and 3 bar.

## 4. Supersymmetry models with MIMAC-He3

We have shown, in the last years, the complementarities of the axial interaction searches with those using mainly the scalar interaction [11].
These complementarities arise from the fact that the relative position of the different supersymmetric models with respect to the exclusion curves are very different when we plot the axial or scalar cross-section as a function of the neutralinos mass. On fig.5 (left) we show the models generated by the DarkSUSY code [14] accessible to MIMAC-He3 and on fig.5 (right) the same models in the scalar cross-section plot. The models are distributed over a wider range of cross-section, and some of them present an extremely small scalar cross-section. If the neutralino nature is described by these models it will be very difficult to detect them by scalar interaction.

A particular interesting point recently addressed in the frame of non universal gaugino masses [12,13] is that neutralino lighter masses than the LEPII limit could be allowed. In such scenarios neutralinos masses of ~10 GeV are proposed requiring an experimental challenge concerning the direct detection threshold. The fact that $^3$He has a light mass allows us to expect to have substantial signal over the threshold of 500 eV.

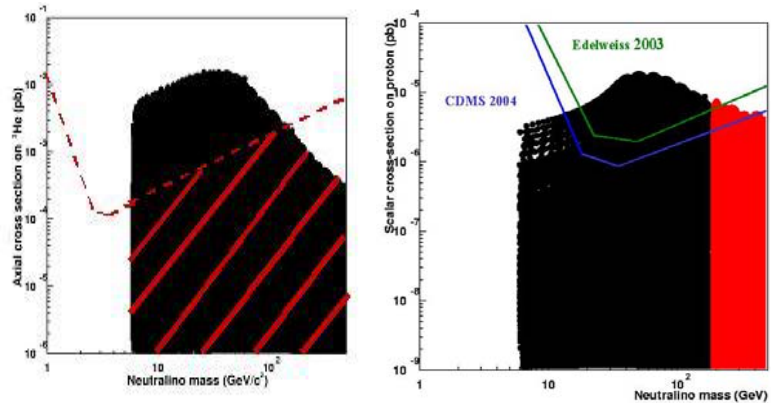

Figure 5 (left): Axial cross-section on $^3$He versus the neutralino mass. The black points correspond to SUSY models allowed by collider as well as cosmological constraints. The projected exclusion curve is displayed for a $10^{-2}$ day$^{-1}$ background level in a 10 kg $^3$He based detector. A lot of models are above the exclusion curve. (right): Scalar cross-section on proton versus the neutralino mass. The black points are the models giving a rate in MIMAC-He3 higher than the $10^{-2}$ day$^{-1}$ background level. The red points are the models giving a rate lower than $10^{-2}$ day$^{-1}$. The exclusion curves from Edelweiss(2003) and CDMS(2004) are shown as a reference. A lot of models accessible to MIMAC-He3, are far below the scalar detection prospects.